\documentclass[]{aa}  
\usepackage{multirow}
\usepackage{graphicx}
\usepackage{hyperref}
\hypersetup
{
    colorlinks= true,
    linkcolor = blue,
    filecolor = blue,      
    urlcolor  = blue,
    citecolor = blue
}
\bibpunct{(}{)}{;}{a}{}{,} 
\usepackage{txfonts}
\usepackage{multirow}
\usepackage{cellspace}
\usepackage[table]{xcolor}

\begin{document}

   \title{Star-formation-rate estimates from water emission}

   \author{K. M. Dutkowska
          %\inst{1}
          \and
          L. E. Kristensen
          }

   \institute{Niels Bohr Institute, Copenhagen University,
              \O{}ster Voldgade 5-7, 1350 Copenhagen K, Denmark\\
              \email{dutkowska@nbi.ku.dk}
             }

   \date{Received xxx xx, xxxx; accepted xxx xx, xxxx}

% \abstract{}{}{}{}{} 
% 5 {} token are mandatory
 
  \abstract
   {The star-formation rate (SFR) quantitatively describes the star-formation process in galaxies throughout cosmic history. Current ways to calibrate this rate do not
usually employ observational methods accounting for the low-mass end of stellar populations as their signatures are too weak.}
   {Accessing the bulk of protostellar activity within  galactic star-forming regions can be achieved by tracing signposts of ongoing star formation. One such signpost is molecular outflows, which are particularly strong at the earliest stages of star formation. These outflows are bright in molecular emission, which is readily observable. We propose to utilize the protostellar outflow emission as a tracer of the SFR.}
   {In this work, we introduce a novel version of the galaxy-in-a-box model, which can be used to relate molecular emission from star formation in galaxies with the SFR. We measured the predicted para-water emission at 988 GHz (which is particularly bright in outflows) and corresponding SFRs for galaxies with $L_\mathrm{FIR} = 10^8 - 10^{11} \mathrm{L}_\odot$ in a distance-independent manner, and compared them with expectations from observations.}
   {We evaluated the derived results by varying star-forming parameters, namely the star formation efficiency, the free-fall time scaling factor, and the initial mass function. We observe that for the chosen water transition, relying on the current Galactic observations and star formation properties, we are underestimating the total galactic emission, while overestimating the SFRs, particularly for more starburst-like configurations.}
   {The current version of the galaxy-in-a-box model only accounts for a limited number of processes and configurations, that is, it focuses on ongoing star formation in massive young clusters in a spiral galaxy. Therefore, the inferred results, which underestimate the emission and overestimate the SFR, are not surprising: known sources of emission are not included in the model. To improve the results, the next version of the model needs to include a more detailed treatment of the entire galactic ecosystem and other processes that would contribute to the emission. Thus, the galaxy-in-a-box model is a promising step toward unveiling the star-forming properties of galaxies across cosmic time.}

   \keywords{Stars: formation --
                Stars: protostars --
                ISM: jets and outflows --
                Galaxies: star clusters: general --
                Galaxies: star formation
               }

   \maketitle

\section{Introduction}

Star formation lies at the very center of the baryon cycle and plays a pivotal role in shaping galactic ecosystems. There are different measures of this process, all of which help to understand and characterize its behavior through cosmic history. One example is the star-formation rate (SFR), as it provides a quantitative description of the star-forming properties of a given object by relating the total mass of stars formed in a give time unit, that is, $M_*/\Delta t$. The SFR is used to establish the cosmic star formation history \citep[e.g.,][]{lilly2013,madau2014}, which in turn is used to understand and quantify the evolution of galaxies. 

The key epoch of cosmic star formation history, which is when star formation peaked  (known as cosmic noon), marks a critical stage during the evolution of today’s galaxy population \citep[e.g.,][]{shapley2011,madau2014,schreiber2020}. Cosmic-noon galaxies, lying at redshifts of 2--3, exhibit extremely different SFRs  from those observed in the local Universe, reaching $>1000\,\mathrm{M}_\odot\,\mathrm{yr}^{-1}$, while the Milky Way is forming stars at a rate of $\sim1\,\mathrm{M}_\odot\,\mathrm{yr}^{-1}$ \citep[e.g.,][]{kennicutt2012}. 

There are various ways of deriving the SFRs in galaxies from nebular line, UV, infrared, radio, and X-ray emission \citep{madau2014}. These methods all assume that there is a scaling between the luminosity in a given band and the SFR. However, the observed emission is usually dominated by high-mass stars, which easily outshine low-mass stars due to their energetic output, and so an initial mass function is applied to correct for low-mass stars, which is where most of the mass resides. In the local Universe, the SFR is readily traced and calibrated with H$\alpha$, H$\beta$, [\ion{O}{ii}], and [\ion{O}{iii}] emission \citep[e.g.,][]{kennicutt1998,tresse2002,kewley2004,salim2007,villavelez2021}. However, in the past 20 years, advances in astrochemistry have provided additional ways to trace star formation, even in its most embedded stages, and allow us to trace the episodes of current star formation in galaxies \citep[e.g.,][]{herbst2009,jorgensen2020}. 

Molecular emission from protostars is not yet commonly used as a SFR tracer. Nevertheless, this emission has the potential to trace even low-mass populations directly. At the earliest stages, the forming star itself is deeply embedded in gas and dust and is thus completely obscured. Therefore, the key is to trace signposts of these early stages that are not obscured. One of these signposts is outflows, which are launched from protostars in their main accretion
phase when the interaction between the infalling envelope, winds,
and jets launched from the protostar are particularly strong \citep{bally2016}. These outflows are launched from close to the protostar, but quickly punch their way through to the surrounding molecular cloud, where they are not obscured \citep{bachiller1990}. In our Galaxy, one of the best tracers of this protostellar component is water \citep{vandishoeck2021}, which is predominantly locked up as ice on dust grains, but is released from the grain mantles into the gas phase, causing a jump in the abundance of many orders of magnitude. At the same time, the physical conditions are conducive to water being readily excited into rotational states \citep[e.g.,][]{suutarinen2014}. 

Water emission is also observed toward high-redshift galaxies \citep[e.g.,][]{yang2013,yang2016,jarugula2019}, where it too has been calibrated to serve as an SFR tracer \citep{jarugula2019}. However, at  high redshift, water is thought to trace dusty molecular clouds illuminated by either massive stars or a central galactic nucleus, and therefore the excitation is assumed to be via far-infrared (FIR) pumping \citep[e.g.,][]{gonzalezalfonso2008,gonzalezalfonso2014}. However, toward the Galactic sources, which were extensively observed with the \textit{Herschel} Space Observatory \citep[e.g., the Water In Star-forming regions with Herschel survey (WISH); ][]{vandishoeck2011,vandishoeck2021}, water emission is almost uniquely associated with outflows, where its excitation is collisionally dominated, and other processes, such as FIR pumping, have a negligible contribution to the excitation \citep{mottram2014, goicoechea2015}.

With the goal of tracing active star formation in galaxies with molecular emission from protostars,  \cite{dutkowska2022} created a galactic model, the so-called galaxy-in-box model, simulating emission from star-forming regions. Using up-to-date knowledge of Galactic star formation and state-of-the-art astrochemical observations of Galactic protostars, the galaxy-in-a-box model simulates emission from young clusters in a chosen galaxy, and at the same time provides insight into the statistics of the star formation process. The default molecular emission is that from water at 988 GHz ($J_{K{\rm a},K{\rm c}} = 2_{02} - 1_{11}$), which is readily observed even at high redshifts where its emission is thought to be dominated by the FIR-pumping-dominated regions, as outlined above.

In this work, we present an extension to the galaxy-in-a-box model, which allows us to derive SFRs from simulated galaxies and their individual star-forming clusters, and to put constraints on local and global SFRs. We focus on water emission at 988 GHz, and simulate emission for galaxies with $L_\mathrm{FIR} = 10^8 - 10^{11} \mathrm{L}_\odot$ for varying star-formation parameters.

This paper is organized as follows. Section \ref{sect:methods} describes all of the changes introduced to the galaxy-in-a-box model. Subsequently, in Section \ref{sect:results} we present the results of this study and test them against observations and the literature; we then discuss these comparisons in Section
\ref{sect:discussion}. Finally, we present our conclusions in Section \ref{sect:conclusions}.

\section{Methods}
\label{sect:methods}

In this study, we explore the relation between the SFR, water luminosity ($L_{\mathrm{H}_2\mathrm{O}}$), and far-infrared luminosity ($L_\mathrm{FIR}$) using the galaxy-in-a-box model \citep[][for an overview of the model see Appendix \ref{app:giab-overview}]{dutkowska2022}. This is a novel, state-of-the-art astrophysical modeling tool that simulates emission from young clusters in a galaxy and provides detailed insights into the constituents of the star-formation process and derived parameters. The model relies on relatively few input parameters, giving the user great flexibility to define global and local galactic parameters. 

For deriving the SFR and relating $L_\mathrm{FIR}$ to the virial mass of galaxies, we implemented a number of upgrades to the model, which we describe in Sect. \ref{subsect:model}. In Sect. \ref{subsect:parameters} we describe the choice of parameters for the simulated galaxies.

\subsection{Changes to the galaxy-in-a-box model}
\label{subsect:model}

For the purposes of this study, we introduced the SFR as an input and output parameter in the galaxy-in-a-box model. We only used the output SFRs. However, in the following, we describe the full extent of the new SFR feature. The SFR tells us how much material is turned into stars per unit of time. With that in mind, we defined the SFR for a cloud in a galaxy as
\begin{equation}
\label{eq:SFR}
\begin{aligned}
\mathrm{SFR_\mathrm{cloud}} = {} & N_* \left( \dfrac{ \langle M_* \rangle}{\mathrm{M}_\odot} \right)
                    \left(  \dfrac{t_\mathrm{cloud}}{\mathrm{yr}}\right)^{-1} \\ 
             = {} & N_* \left( \dfrac{ \langle M_* \rangle }{\mathrm{M}_\odot} \right) 
                    \left(  \dfrac{\tau^\mathrm{sc}_\mathrm{ff} t_\mathrm{ff}}{\mathrm{yr}}\right)^{-1} ,\
\end{aligned}
\end{equation}
where $N_*$ is the number of formed protostars, $\langle M_* \rangle$ is the average protostellar mass, $t_\mathrm{cloud}$ is the age of the cloud, $\tau^\mathrm{sc}_\mathrm{ff}$ is the unitless free-fall scaling factor \citep{dutkowska2022}, and $t_\mathrm{ff}$ is the free-fall time of the cloud. In the case of the galaxy-in-a-box model, the age is randomized, that is, it randomly scales the ages such that they range from newly formed to completely collapsed. The global SFR of the entire galaxy is therefore the sum of the individual rates for each cloud.

In the model, we assume that each cloud goes on to form one cluster; in nature, clouds may go on to form several generations of clusters, but for the purposes of this study, where we consider global star formation, this is not relevant. With this implementation of the SFR, we introduce the possibility to also constrain the SFR at the cloud or cluster level. The cluster module can now be run with a fixed SFR, where the age of the cluster is adjusted through the free-fall time scaling factor, which can be easily derived from Eq. (\ref{eq:SFR}):

\begin{equation}
\label{eq:sfr_scaling}
    \tau^\mathrm{sc}_\mathrm{ff} = 
    N_* \left( \dfrac{\langle M_* \rangle }{\mathrm{M}_\odot} \right)
    \left( \dfrac{t_\mathrm{ff, random}}{\mathrm{Myr}}\right)
    \left( \dfrac{\mathrm{SFR}_\mathrm{cloud}}{\mathrm{M}_\odot\,\mathrm{Myr}^{-1}}\right) .\
\end{equation}
In this equation, $t_\mathrm{ff}$ is already randomized ($t_\mathrm{ff, random}$) to avoid poor SFR adjustment due to age assignment that takes place later in the model. However, Eq. (\ref{eq:sfr_scaling}) is not used in this study. 

On a global scale, that is, when introducing constraints on the total galactic SFR, the new version of the galaxy-in-a-box model monitors the total SFR of the given galaxy and computations stop when the specified SFR is reached. The allowed deviation from the specified SFR is $\pm10\%$.  There may be situations where the galaxy-in-a-box model will not converge: these situations are unphysical, and an example would be a very low-mass galaxy with a very high SFR. There needs to be enough gas that can be turned into stars at the desired rate. 

One of the changes to the galaxy-in-a-box model that was essential for this study was to set the number of clusters as limited by the total molecular gas reservoir, rather than setting it as a fixed number in the input file. We infer the number of molecular clouds from which star forming clusters form by putting an upper limit on the total mass of the molecular clouds, which are randomly generated using the molecular cloud mass distribution. When the limit is reached, the clouds are no longer passed to the cluster part of the calculations \citep[see Fig. 1 of][]{dutkowska2022}. This way we ensure that the mass of clouds does not exceed the available molecular reservoir.

\begin{figure}[t!]
\resizebox{\hsize}{!}{\includegraphics{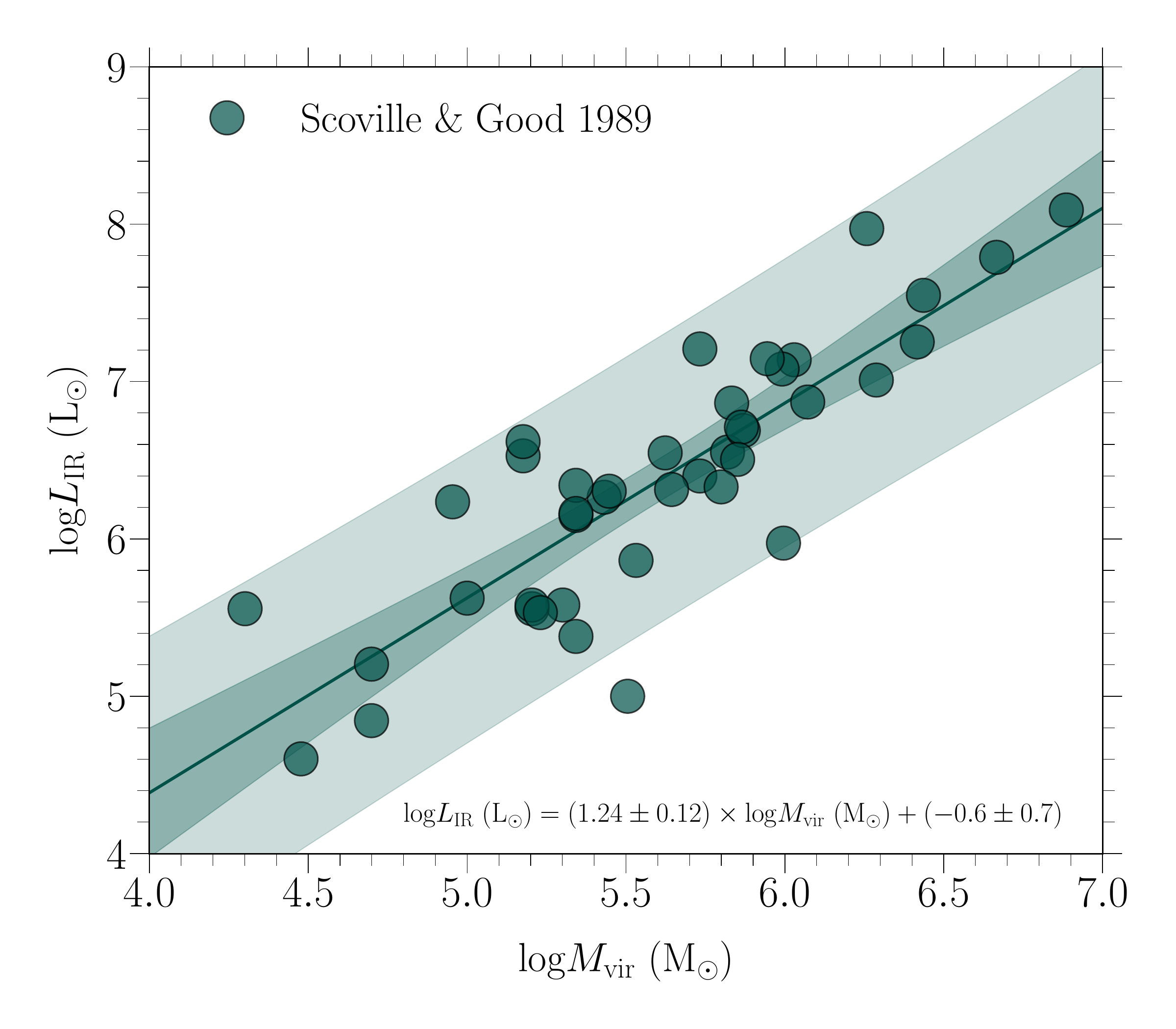}}
\caption{Correlation between log $M_\mathrm{vir}$ and log $\,L_\mathrm{FIR}$ following \cite{scoville1989}. The solid straight line represents the best-fit power law to the data points, the darker shaded region corresponds to the 95\% confidence region of the correlation, and the lighter shaded region represents the region that contains 95\% of the measurements.}
\label{fig:scoville}
\end{figure}

Lastly, the mass properties of the galaxy can now be set by defining the $L_\mathrm{FIR}$ of the galaxy. Following \cite{scoville1989}, the model derives the mass of the molecular reservoir through the observed $M_\mathrm{vir}-L_\mathrm{IR}$ relation (see Fig. \ref{fig:scoville}). The viral mass of the galaxy can be expressed as:

\begin{equation}
\label{eq:lum-to-mass}
        \dfrac{M_\mathrm{vir}}{\mathrm{M}_\odot} = 10^{0.5\pm0.6}\left( \dfrac{L_\mathrm{IR}}{\mathrm{L}_\odot}\right)^{0.81\pm0.08} ,\
\end{equation} 
where $L_\mathrm{IR}$ is the total far-infrared luminosity of the cloud. With Eq. (\ref{eq:lum-to-mass}), we can simulate $L_{\mathrm{H}_2\mathrm{O}}$ for galaxies with different $L_\mathrm{IR}$, including typical galaxy types observed with H$_2$O emission, that is, subluminous infrared galaxies (subLIRGs; $ L_\mathrm{IR} <  10^{11}\ \mathrm{L}_\odot$),  LIRGs $(10^{11}\ \mathrm{L}_\odot \leq L_\mathrm{IR} < 10^{12}\ \mathrm{L}_\odot$), ultraluminous infrared galaxies (ULIRGs; $10^{12}\ \mathrm{L}_\odot \leq L_\mathrm{IR} < 10^{13}\ \mathrm{L}_\odot$), and hyperluminous infrared galaxies (HyLIRGs; $L_\mathrm{IR} \geq 10^{13}\ \mathrm{L}_\odot$). In this study, we are interested in relative values for derived luminosities and SFRs, and therefore we make an assumption that $L_\mathrm{FIR}$ is a proxy for $L_\mathrm{IR}$, and use these luminosities interchangeably. 

\subsection{Considered parameters}
\label{subsect:parameters}

The goal of this study is to explore the SFRs derived with the galaxy-in-a-box model and how they relate to derived luminosities. To achieve this goal, we decided to use the template galaxy from \cite{dutkowska2022} with emission from the para-H$_2$O \(2_{02} - 1_{11}\) line at 987.927 GHz, and tweak the star formation efficiency, the free-fall-time scaling factor, and the initial mass function. The exact ranges of the probed parameters are described in Table \ref{tab:params}.

\begin{table}[t]
\caption{Parameters considered in this study}
\label{tab:params} 
\centering
\begin{tabular}{lcc|cccc}
\hline \hline
\multicolumn{3}{l|}{\multirow{2}{*}{}}  & \multicolumn{4}{c}{Galactic type $\left( \log(L_\mathrm{FIR}/L_\odot) \right)$} \\ \cline{4-7} 
\multicolumn{3}{l|}{}                   & $8-8.9$ 
          & $9-9.9$ 
          & $10-10.9$   
          & 11  \\ \hline
\multicolumn{1}{l|}{\multirow{8}{*}{\rotatebox[]{90}{Parameter}}} & \multicolumn{1}{c|}{\multirow{3}{*}{$\varepsilon_\mathrm{SF}$}} 
                                                               & $1\%$  & \cellcolor{red!20}x & \cellcolor{red!20}x & \cellcolor{red!20}x & \cellcolor{red!20}x \\
\multicolumn{1}{l|}{} & \multicolumn{1}{c|}{}                  & $10\%$ & \cellcolor{red!20}x & \cellcolor{red!20}x & \cellcolor{red!20}x & \cellcolor{red!20}x \\
\multicolumn{1}{l|}{} & \multicolumn{1}{c|}{}                  & $30\%$ &          x          &          x          &                     &          x          \\
\multicolumn{1}{l|}{} & \multicolumn{1}{c|}{\multirow{2}{*}{$\tau_\mathrm{ff}^\mathrm{sc}$}} & 1 & \cellcolor{red!20}x & \cellcolor{red!20}x & \cellcolor{red!20}x & \cellcolor{red!20}x \\
\multicolumn{1}{l|}{} & \multicolumn{1}{c|}{}                  & 5 & x  &  x   &  x   &  x  \\
\multicolumn{1}{l|}{} & \multicolumn{1}{c|}{\multirow{3}{*}{IMF}} & $s$   & \cellcolor{red!20}x & \cellcolor{red!20}x & \cellcolor{red!20}x & \cellcolor{red!20}x \\
\multicolumn{1}{l|}{} & \multicolumn{1}{c|}{}                     & $t$-$h$ & \cellcolor{red!20}x & \cellcolor{red!20}x & \cellcolor{red!20}x & \cellcolor{red!20}x \\
\multicolumn{1}{l|}{} & \multicolumn{1}{c|}{}                     & $b$-$h$ &          x          &          x          &          x          &          x          \\
\hline
\end{tabular}
\tablefoot{Star-forming and galactic parameters considered in this study. Red filling corresponds to parameter space used for the high-$z$ correlation test, while x refers to those considered together with the standard correlation. For the latter, the only omitted galactic type was that with $L_\mathrm{FIR} = 10^{10} \mathrm{L}_\odot$ for all combinations involving $\varepsilon_\mathrm{SF} = 30\%$. In the IMF parameters, \textit{s} refers to standard, \textit{t-h} to top-heavy, and \textit{b-h} to bottom-heavy.}
\end{table}

For the galactic masses, or in this case luminosities, we decided to probe galaxies with $L_\mathrm{FIR} = 10^8 - 10^{11}\ \mathrm{L}_\odot$, where for the range $10^8 - 10^{10}\ \mathrm{L}_\odot$ we continued with an increment corresponding to the given order of magnitude (i.e., $10^8, 2\times10^8, 3\times10^8$, etc.), and we stopped at $10^{11}\ \mathrm{L}_\odot$. We made this choice because we wanted to probe the chosen regime in a relatively uniform way. Moreover, the lower limit was dictated by low galactic mass ($10^8\ \mathrm{L}_\odot$ corresponds to $\sim10^7\ \mathrm{M}_\odot$), while the upper one was dictated by the limitations of the computational power. As shown below, the inferred SFRs can readily be extrapolated to even higher luminosities. 

In the model, we use the relation between mass and H$_2$O line luminosity obtained only from Galactic sources to estimate the amount of emission generated by protostars. As a sanity check, we can include the high-$z$ observations in this correlation, as shown in Fig. \ref{fig:gal_highz}. Including the high-$z$ measurements shifts the correlation slightly, such that low-mass protostars are assigned less emission, and vice versa for high-mass protostars. Therefore, if these high-$z$  sources are included, this has implications for the assumed IMF. In order to obtain luminosity distances for high-$z$ objects, we used a Planck
2018 flat $\Lambda\mathrm{CDM}$ cosmology with $H_0 = 67.7\;\mathrm{km\,s}^{-1}\,\mathrm{Mpc}^{-1}$ and $\Omega_M = 0.310,$ as implemented in the \textsc{Astropy} package \citep{astropy2018}.

\section{Results}
\label{sect:results}

By extracting SFRs together with $L_{\mathrm{H}_2\mathrm{O}}$, while at the same time defining galaxies according to their luminosity  rather than their mass directly, we are able to confront expectations based on the literature about the star formation process as seen in the Milky Way while simultaneously testing the galaxy-in-a-box model. Therefore, in this proof-of-concept study, we ran a number of simulations spanning a range of parameters representing different galactic and star formation properties (see Table \ref{tab:params}). 

As mentioned in Sect. \ref{subsect:model}, we used two different mass--line luminosity correlations: in the first, we only use the Galactic data points, and in the second we include the high-$z$ data points. We excluded certain parameters from the high-$z$ test, because they were either computationally heavy or unnecessary for testing the impact of the high-$z$ extrapolation (for further discussion, see Sect. \ref{sect:discussion}). For galaxies with $L_\mathrm{FIR} = 10^8 - 9\times10^8\ \mathrm{L}_\odot$, we ran 40 simulations for each setup, while for other luminosity ranges we ran 20 simulations per setup. The increased number of simulations for this specific galactic type was dictated by higher SFR variations, as the molecular reservoir is relatively low, which is reflected in larger variations in the number of formed stars. We also excluded calculations for galaxies with $L_\mathrm{FIR} = 10^{10} - 9\times10^{10}\ \mathrm{L}_\odot$, which would have $\varepsilon_\mathrm{SF} = 30\%$, because they were the most computationally heavy, and including them would not affect any conclusions of this study. In total, we ran 15200 simulations, including 12240 main runs and 2960 runs for the high-$z$ test.

Uncertainties for the simulation results are calculated as a standard deviation from the mean value, which is derived for all runs with the same set of parameters. The best fits were obtained using linear regression while accounting for the spread in the y-direction. If the spread is not shown, this means that the size is smaller than the marker or the line size. When recalculating fluxes to luminosities, we naturally account for the propagation of uncertainties. 

We describe the literature sample chosen for this study in Sect. \ref{subsect:literature}. We then present the results through derived $L_\mathrm{FIR}\,-\,L_{\mathrm{H}_2\mathrm{O}}$ (Sect. \ref{subsect:lfir_lh2o}), $L_\mathrm{FIR}\,-\,$SFR (Sect. \ref{subsect:lfir_sfr}), and $L_{\mathrm{H}_2\mathrm{O}}\,-\,$SFR (Sect. \ref{subsect:lh2o_sfr}) relations, which we compare to those provided in the literature.

\subsection{Literature sample}
\label{subsect:literature}

As a default source of Galactic observations, we use data from the Water Emission Database \citep{dutkowska2022} for the para-H$_2$O \(2_{02} - 1_{11}\) line at 987.927 GHz, which consists of Galactic low-, intermediate-, and high-mass protostars observed as part of WISH \citep{vandishoeck2011} and the William Herschel Line Legacy Survey \citep[WILL;][]{mottram2017}. 

The sample of extragalactic sources used in the high-$z$ test was taken directly from \cite{werf2011}, \cite{combes2012}, \cite{omont2013}, \cite{riechers2013}, \cite{yang2013}, \cite{yang2016}, \cite{apostolovski2019}, and \cite{jarugula2019}. This sample includes both nearby subLIRGs, LIRGs, and quasars, as well as high-z quasars, ULIRGS, and HyLIRGs, with the farthest one being the HyLIRG, namely HFLS3,  at $z = 6.337$ \citep[$D_\mathrm{L} = 62834.75$ Mpc; for more details see][]{riechers2013}. A detailed description of the sample and exact values used in this study can be found in \citet{kristensen2022sub}.

\begin{figure}[]
\resizebox{\hsize}{!}{\includegraphics{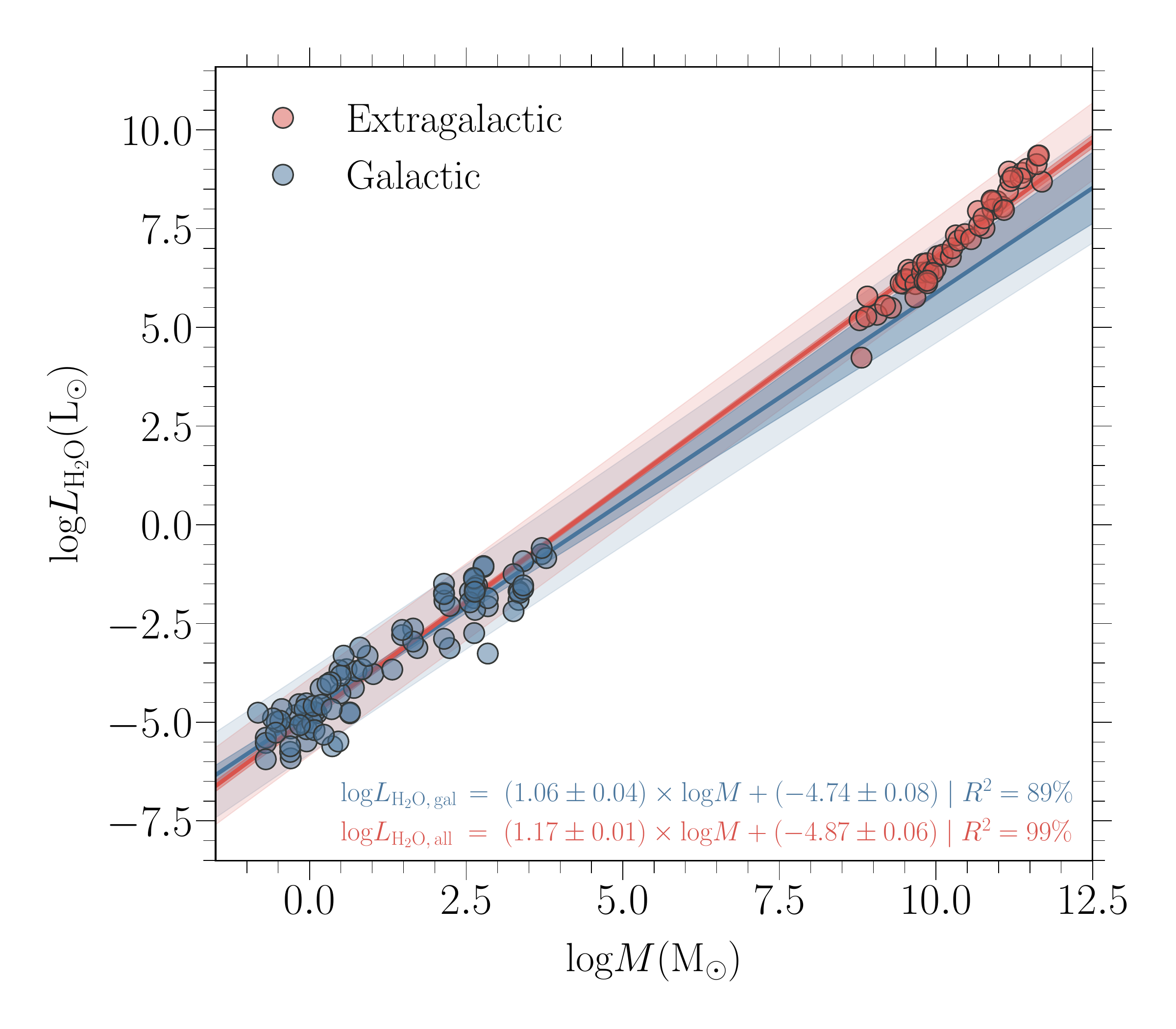}}
\caption{Two types of correlations and observational samples used in this study. The blue solid line corresponds to the best fit to the Galactic with the data points taken from the Water Emission Database \citep{dutkowska2022}, while the red solid line represent the best fit that also includes the extragalactic sample consisting of nearby subLIRGs, LIRGs, and quasars, as well as high-$z$ quasars, ULIRGs, and HyLIRGs (for details see Sect. \ref{subsect:literature}). Markers correspond to the observations from each sample. Shading follows that from Fig. \ref{fig:scoville}.}
\label{fig:gal_highz}
\end{figure}

\subsection{\texorpdfstring{Total stellar mass versus SFR}{M* vs SFR}}
\label{subsect:mass_sfr}

\begin{figure*}[]
\resizebox{\hsize}{!}{\includegraphics[width=\textwidth]{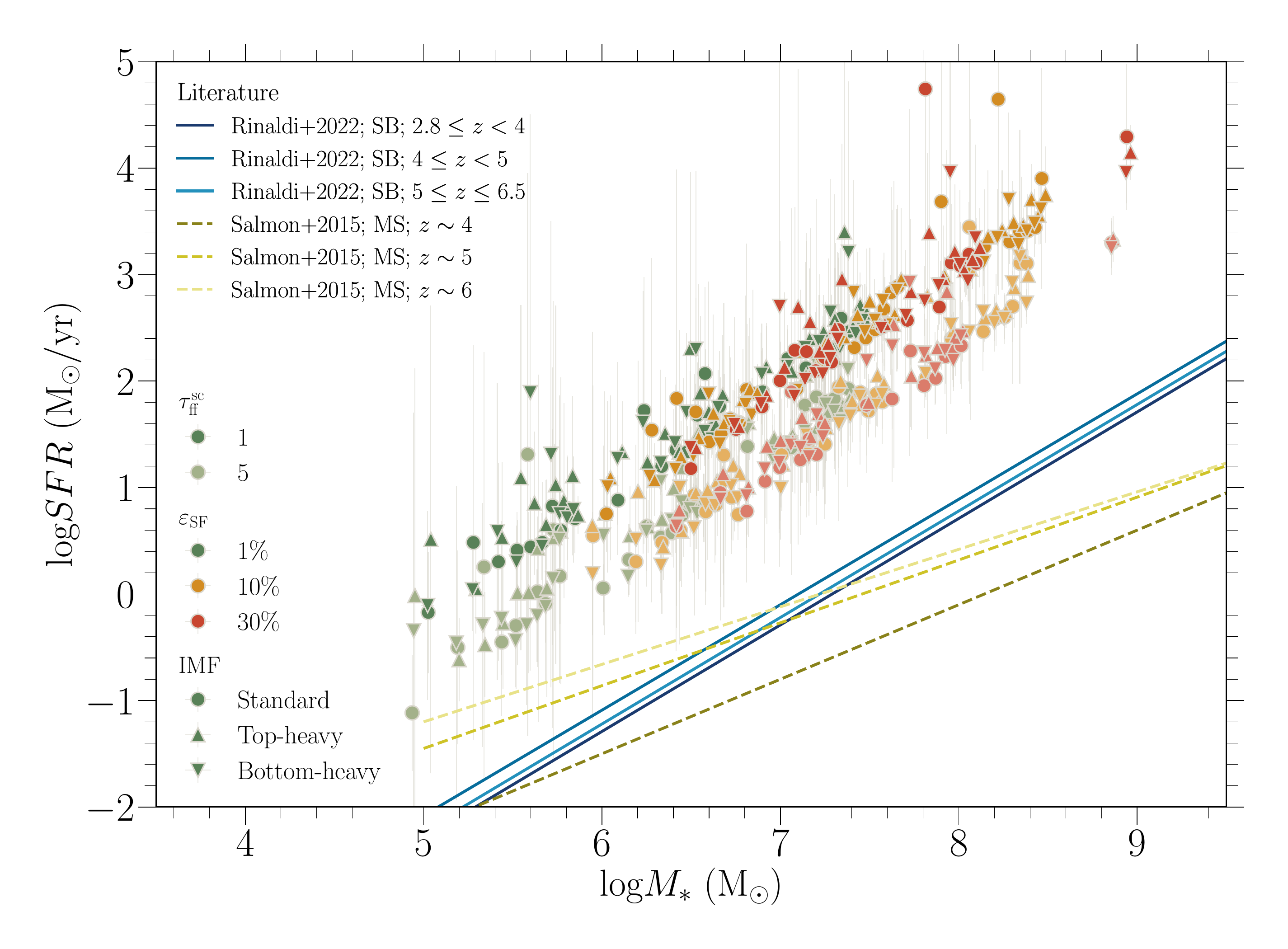}}
\caption{SFR as a function of stellar mass of each galaxy. Full color markers represent results, where the free-fall-time scaling factor was set to 1, while markers in the same but lighter colors correspond to $\tau^\mathrm{sc}_\mathrm{ff}$ of 5. Circles represent setups with the standard IMF \citep{chabrier2003}, while triangles pointing upwards and downwards represent setups with its top-heavy and bottom-heavy versions, respectively. Different colors of the markers refer to different star formation efficiencies, where green, orange, and red mean an $\varepsilon_\mathrm{SF}$ of 1\%, 10\%, and 30\%, respectively. Solid lines represent best-fit lines from \cite{rinaldi2022} to their starburst (SB) population, while dashed lines represent best fits to the main sequence galaxies from \cite{salmon2015}.}
\label{fig:mass_sfr}
\end{figure*}

We evaluated the derived SFRs by exploring their relation with the total stellar mass of the corresponding galaxies. From  Fig. \ref{fig:mass_sfr}, we see that we are overestimating the SFRs when looking at functions derived by for example \cite{salmon2015} for the main sequence galaxies and \cite{rinaldi2022} for the starbursts. 

With the chosen set of properties, galaxies with $M_*<10^{6.5}\;\mathrm{M}_\odot$ seem to lie close to the main sequence estimates from \cite{salmon2015}, at least in their lower limits. However, going to cases where the combination of considered parameters resulted in an increase in SFRs, especially galaxies with $M_* > 10^{6.5}\;\mathrm{M}_\odot$, we start overestimating SFRs by at least one order of magnitude when compared to the literature \citep{rinaldi2022}. 

We also observe two distinct populations that appear to be dictated by the value of the free-fall-time scaling factor. For $\tau^\mathrm{sc}_\mathrm{ff} = 1$, we let the efficiency of the free-fall time depend only on the density of the progenitor molecular cloud, while by introducing $\tau^\mathrm{sc}_\mathrm{ff} = 5$ we prolong the time required to form most of the stellar population, resulting in a more diverse range of protostellar ages. From Fig. \ref{fig:mass_sfr}, we see how a decrease in  the free-fall-time scaling factor influences the derived SFR. Considering the relatively low efficiency of the star formation process, the lower-SFR branch is likely to be more consistent with the nature of the star formation process. We discuss this topic further in Sect. \ref{subsect:params}.

\subsection{\texorpdfstring{$L_\mathrm{FIR}\,-\,L_{\mathrm{H}_2\mathrm{O}}$ correlation}{LFIR - LH2O correlation}}
\label{subsect:lfir_lh2o}

To compare the predicted fluxes with observations, we first converted them to luminosities using the following expression:

\begin{equation}
\label{eq:flux_to_lum}
    \dfrac{L_\mathrm{line}}{\mathrm{L}_\odot} = 99.04 \left(\dfrac{ I}{\mathrm{Jy\,km\,s}^{-1}}\right)
    \left(\dfrac{\lambda_0}{\mu\mathrm{m}}\right)^{-1}
    \left(\dfrac{D_\mathrm{L}}{\mathrm{Mpc}} \right)^2 ,\
\end{equation}
where $I$ is the total intensity in Jy~km~s$^{-1}$, $\lambda_0$ the wavelength in microns ($303.4557\,\mu\mathrm{m}$ for the para-H$_2$O \(2_{02} - 1_{11}\) line), and $D_\mathrm{L}$ the luminosity distance of the source in megaparsecs. By converting fluxes, we can quantitatively compare our results with observations, as they are no longer distance dependent. 

Using linear regression, we derived best-fit lines to the following expression:

\begin{equation}
\label{eq:fit}
    \log_{10} \left( L_{\mathrm{H}_2\mathrm{O}} / \mathrm{L}_\odot\right)= a\times\log_{10} \left( L_\mathrm{FIR} / \mathrm{L}_\odot\right) + b .\
\end{equation}
Table \ref{tab:relations} provides all of the derived slopes and intercepts. In the following, we focus on the two setups exhibiting the highest and lowest water emission. These are the models with $\varepsilon_\mathrm{SF}$=30\%, IMF = top-heavy, $\tau^\mathrm{sc}_\mathrm{ff}$=1, and $\varepsilon_\mathrm{SF}$=1\%, IMF = bottom-heavy, $\tau^\mathrm{sc}_\mathrm{ff}$=5, respectively. 
For the least emitting case, we derive $a = 0.809 \pm 0.003$ and $b = -7.269 \pm 0.029$, while for the most emitting case we derive $a = 0.809 \pm 0.001$ and $b = -5.135 \pm 0.012$. In both cases, $R^2 = 99.9\%$. For all of the simulations, the slope stays roughly constant with $a \approx 0.81$, and therefore the span of luminosities is described by the intercept falling in the range of $-$7.269 to $-$5.135. We can derive the general relation for water-line luminosity depending on the intercept value:

\begin{equation}
\label{eq:fit2}
    L_{\mathrm{H}_2\mathrm{O}} / \mathrm{L}_\odot  = 10^b \left( L_\mathrm{FIR} / \mathrm{L}_\odot\right)^{0.81} .\
\end{equation}

\begin{figure*}[]
\resizebox{\hsize}{!}{\includegraphics[]{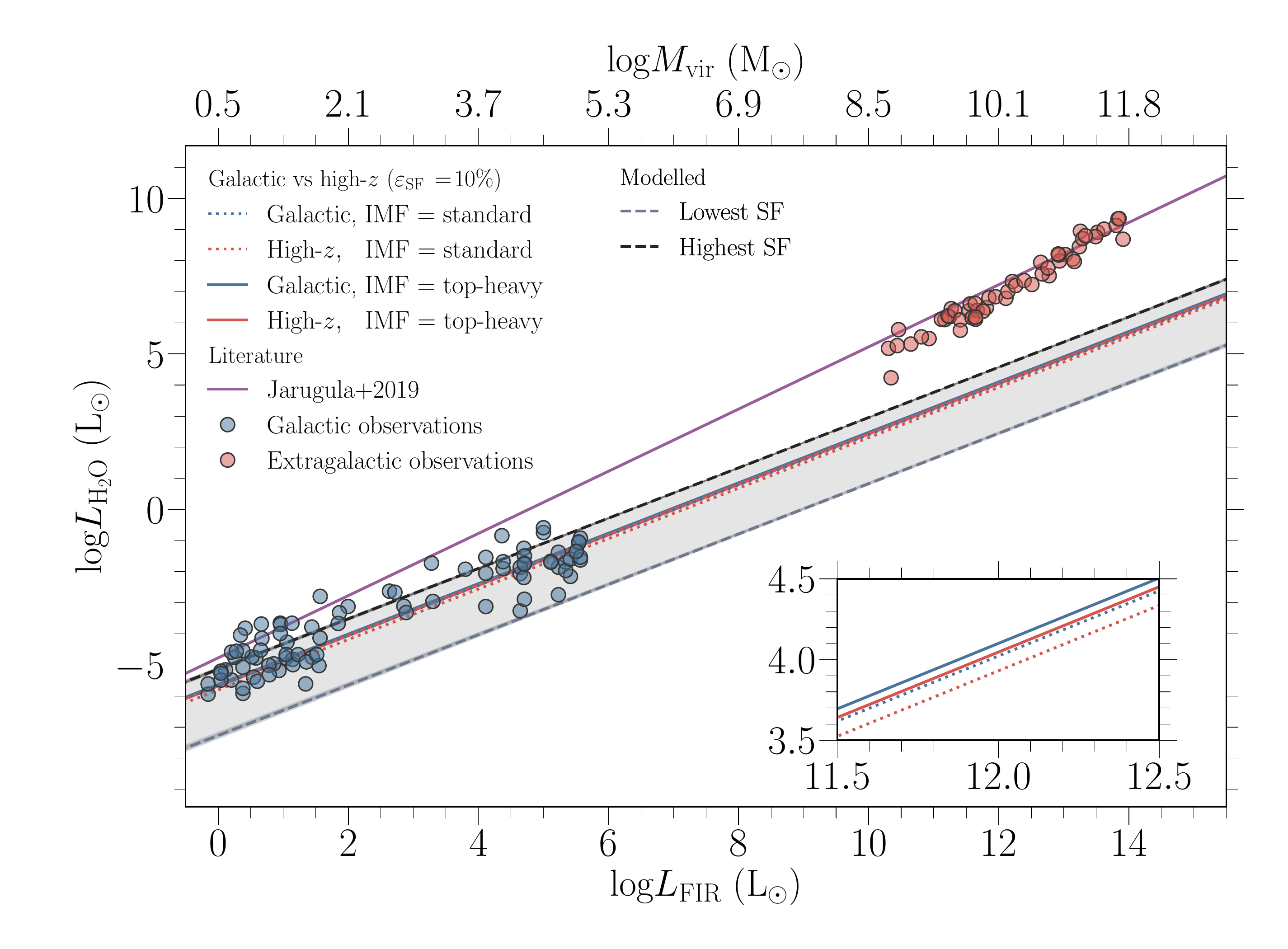}}
\caption{Simulated water-line luminosity as a function of $L_\mathrm{FIR}$. The dashed black line represents the most emitting galaxy in our simulations ($\varepsilon_\mathrm{SF}$=30\%, IMF = top-heavy and $\tau^\mathrm{sc}_\mathrm{ff}$=1), while the dashed gray line corresponds to the least emitting one ($\varepsilon_\mathrm{SF}$=1\%, IMF = bottom-heavy and $\tau^\mathrm{sc}_\mathrm{ff}$=5). The gray-shaded area between these two lines refers to the probed parameter space, and all possible outcomes considered in this study would fall in that regime. Solid blue and red lines refer to the results derived for setups with the top-heavy IMF form for the Galactic and extragalactic $M_\mathrm{env} - I$ relations, respectively. Dotted lines show the results for these two correlations, when the standard IMF is applied. In both, i.e., the standard and the top-heavy cases, the free-fall-time scaling factor is set to 1. Circles refer to observational samples (for more details we refer the reader to Sect. \ref{subsect:literature}), while the purple line represents the expected relation from \cite{jarugula2019}.}
\label{fig:lfir_lh2o}
\end{figure*}

From Fig. \ref{fig:lfir_lh2o}, we see that we deviate from extragalactic observations by between a factor of a few and about two orders of magnitude. We observe that the expectations built on the extragalactic sample taken from \cite{jarugula2019} ---where \(L_{\mathrm{H}_2\mathrm{O}}/L_\mathrm{FIR} = 1.69\substack{+0.79 \\ -0.54}\times10^{-5}\) (we explore this more extensively in Sect. \ref{subsect:ratio})--- are especially far from our expectations for the brightest high-$z$ galaxies. We discuss this further in Sect. \ref{subsect:lfir_lh2o}. Also, in Sect. \ref{subsect:test}, we explore the possible impact of the inclusion of high-$z$ starbursts on the correlation between the envelope mass and intensity ($M_\mathrm{env} - I$ relation) ---which is the basis of emission assignment in the galaxy-in-a-box model--- and whether it could explain the observed differences.

\subsection{\texorpdfstring{$L_\mathrm{FIR}\,-\,$SFR correlation}{LFIR - SFR correlation}}
\label{subsect:lfir_sfr}

\begin{figure}
\resizebox{\hsize}{!}{\includegraphics{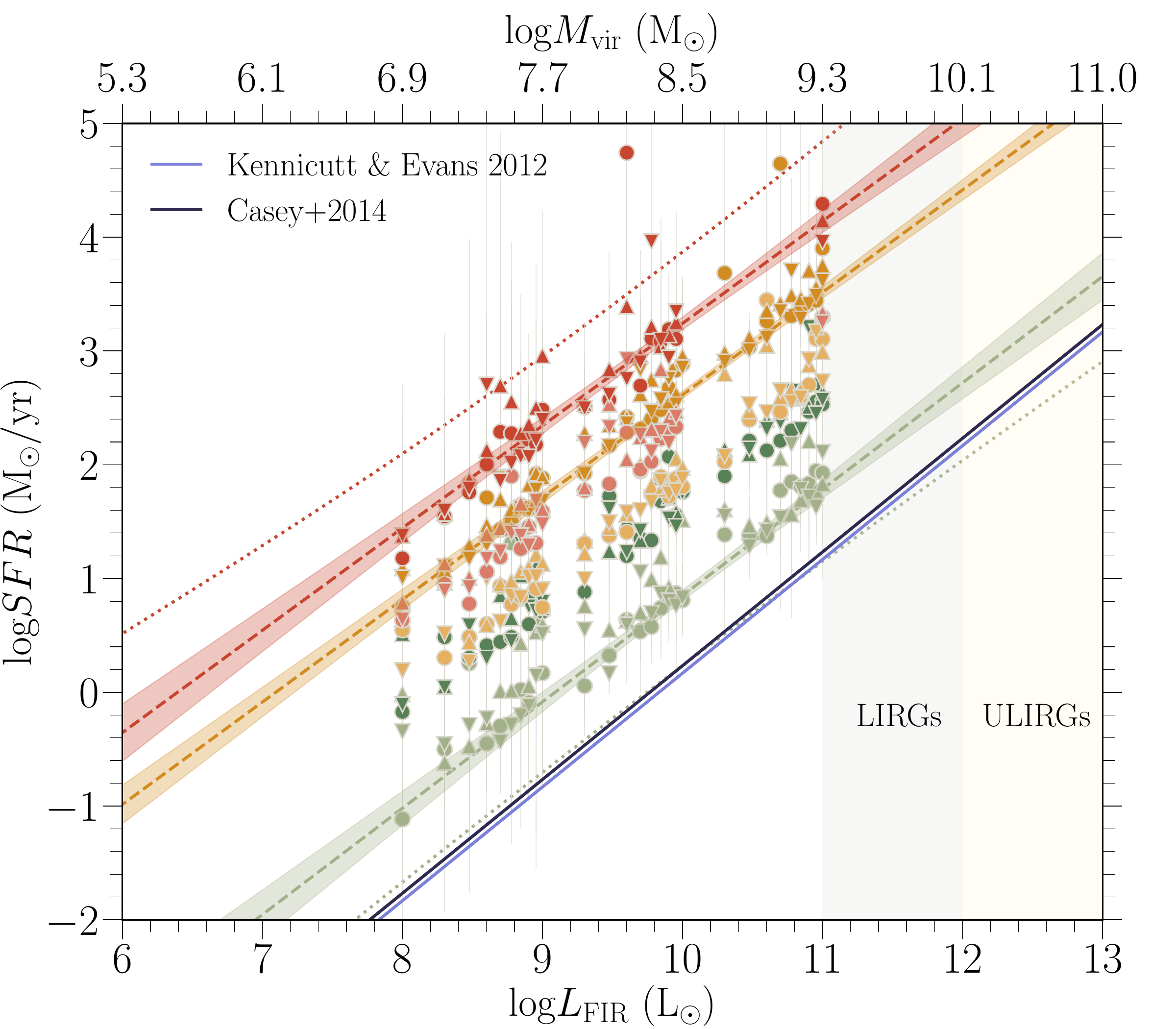}}
\caption{SFR as a function of $L_\mathrm{FIR}$. Colors and markers as in Fig. \ref{fig:mass_sfr}. Dotted lines refer to the upper prediction band for the  setup with the highest SFR and the lower prediction band for the setup with the lowest SFR. Shading of the best-fit lines corresponds to the 95\% confidence region of each correlation. Solid lines represent the literature estimates.}
\label{fig:lfir_sfr}
\end{figure}

To further evaluate derived SFRs, we explored their relation with corresponding far-infrared luminosities (Fig. \ref{fig:lfir_sfr}). We clearly see that the derived SFRs create different populations depending on the star formation efficiency and the free-fall-time scaling factor. Again, we are clearly overestimating the SFRs. However, relations in the literature, for example those of \cite{kennicutt2012} and \cite{casey2014}, fall into our lower prediction regime, meaning that at least for the star forming galaxies with lower star formation activity (with respect to the standard setup in the galaxy-in-a-box model), we are roughly recovering the expected star formation process. 

The span of the SFRs derived in this study depends strongly on the efficiency of the process. The discrepancy between the literature values and our simulations can be as high as two orders of magnitude. We focused on and derived relations analogous to Eq. (\ref{eq:fit}) for the setups with the lowest and highest emission, as well as the standard model setup from the galaxy-in-a-box model. We provide all of the derived relations in Table \ref{tab:relations}. Here, we do not derive almost identical slopes, as we did for $L_\mathrm{FIR}- L_{\mathrm{H}_2\mathrm{O}}$. For the most extreme cases of the $L_\mathrm{FIR}$--SFR relation, we derive slopes of $0.94\pm0.04$ and $0.90\pm0.03$, which agree within the uncertainties, while the derived intercepts (here, the intercept refers to the term $b$ in Eq. (\ref{eq:fit}), which is further used as showed in Eq. (\ref{eq:fit2})) are equal to $-8.50\pm0.35$ and $-5.75\pm0.33$, respectively. We further discuss the apparent excess in SFR in Sect. \ref{fig:lfir_sfr}.

\subsection{\texorpdfstring{$L_{\mathrm{H}_2\mathrm{O}}\,-\,$SFR correlation}{LH2O - SFR correlation}}
\label{subsect:lh2o_sfr}

\begin{figure}[]
\resizebox{\hsize}{!}{\includegraphics{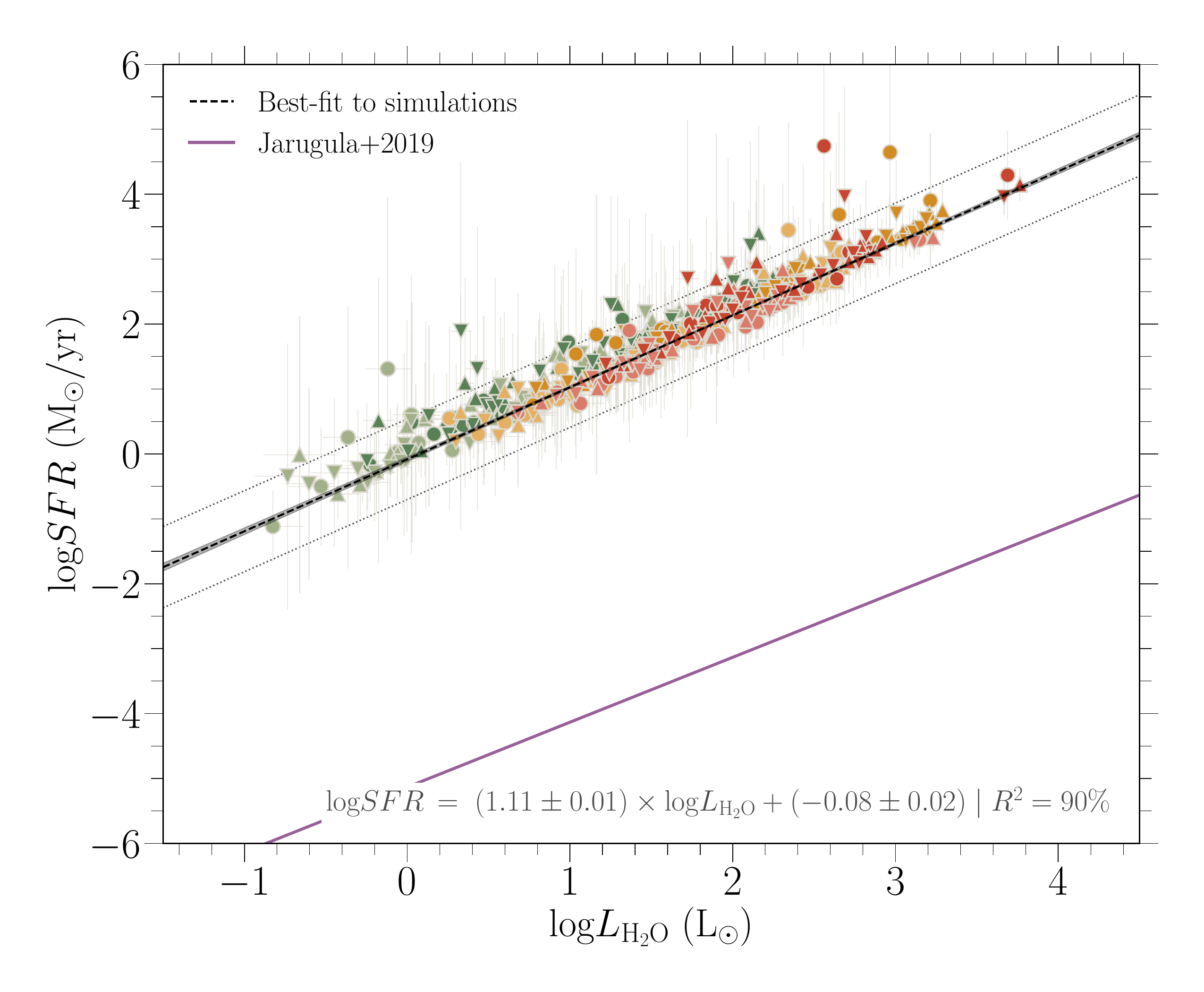}}
\caption{SFR as a function of water line luminosity. Colors and markers are as in Fig. \ref{fig:mass_sfr}. The shaded region corresponds to the 95\% confidence region of the correlation, while dotted lines indicate where 95\% of the measurements should fall. The solid purple line represents the expected relation from \cite{jarugula2019}.}
\label{fig:lumi_sfr}
\end{figure}

The last explored dependence was that of $L_{\mathrm{H}_2\mathrm{O}}$ and the corresponding SFRs. We see from Fig. \ref{fig:lumi_sfr} that all of the derived SFRs fall into the same population, which is expected considering the fact that the greater the luminosity, the more actively star-forming and massive the corresponding galaxy. By fitting all of the derived points to Eq. (\ref{eq:fit}), we get a slope of $1.11\pm0.01$ and an intercept of $-0.083\pm0.018$, indicating a near-proportionality between the SFR and $L_{\mathrm{H}_2\mathrm{O}}$. 

However, Fig. \ref{fig:lumi_sfr} suggests that we are systematically  overestimating SFRs by approximately four orders of magnitude with respect to the findings of \cite{jarugula2019}, where \( \mathrm{SFR}\left(\mathrm{M}_\odot\mathrm{yr}^{-1}\right) = 7.35\substack{+5.74 \\ -3.22}\times10^{-6}L_{\mathrm{H}_2\mathrm{O}}\left( \mathrm{L}_\odot \right)\). If extrapolating their relation to Galactic star-forming regions, we would underestimate SFRs by orders of magnitude \citep{kristensen2022sub}. We discuss this discrepancy in Sect. \ref{subsect:obs}.

\section{Discussion}
\label{sect:discussion}

In the following, we discuss derived SFRs and water luminosities. We also evaluate how the star-formation parameters considered here could affect the results and compare our results with the literature. Moreover, we discuss what other physical processes not considered in this study could impact the derived values and explore other possible influences.

\subsection{\texorpdfstring{Insights from $L_{\mathrm{H}_2\mathrm{O}}/L_\mathrm{FIR}$ ratios}{Insights from LH2O/LFIR ratios}}
\label{subsect:ratio}

\begin{figure}[]
\resizebox{\hsize}{!}{\includegraphics{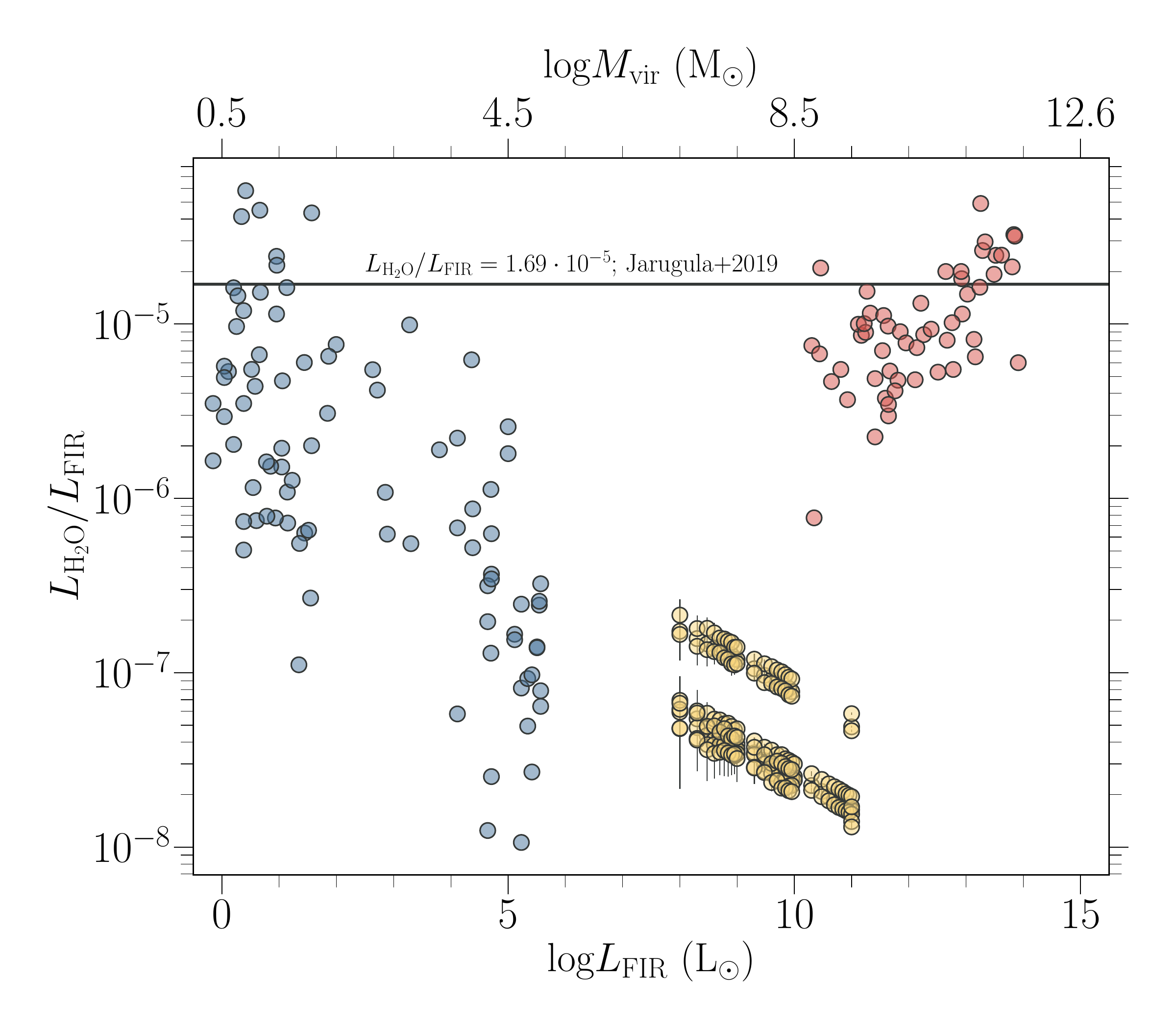}}
\caption{$L_{\mathrm{H}_2\mathrm{O}}/L_\mathrm{FIR}$ as a function of $L_\mathrm{FIR}$. Blue and red points refer to Galactic and extragalactic observations, respectively. Yellow points refer to our simulations with star formation efficiencies of 10\% and 30\%. }
\label{fig:ratio}
\end{figure}

The ratio of $L_{\mathrm{H}_2\mathrm{O}}$ and corresponding $L_\mathrm{FIR}$ could be used to understand the source of the observed water emission (this is shown in Fig. \ref{fig:ratio}). This in turn can help us to understand whether or not water behaves differently in different galactic regions and galactic types. With this in mind, we calculated the ratios derived from the galaxy-in-a-box model and compared them with our Galactic and extragalactic samples. 

The derived values $ 10^{-8} < L_{\mathrm{H}_2\mathrm{O}}/L_\mathrm{FIR} < 10^{-6}$ fall below those from all objects considered in the extragalactic sample, but coincide with the Galactic sample at its high-mass/high-luminosity end (see Fig. \ref{fig:lfir_lh2o}). We know from Galactic observations \citep[e.g.,][]{vandishoeck2021} that water emission from young stellar objects predominantly comes from the shocked material in outflows. Therefore, a natural assumption would be that the Galactic sample is consistent in terms of the calculated ratios. Instead, what we see is that low- to intermediate-mass protostars exhibit roughly the same ratios as the extragalactic sample, and we see a clear drop for the most luminous end of the Galactic objects. 

Available water observations of Galactic high-mass young stellar objects are limited because of both their number and sensitivity. One of the most detailed studies was conducted with a survey towards the Cygnus X star-forming region \citep[PI: Bontemps;][]{phdSJG2015}. Cygnus-X is one of the nearest  massive star-forming complexes \citep[$D$ $\sim$ 1.3--1.4 kpc, e.g.,][]{rygl2012}. However, even these observations do not recover the total emission that would come from a high-mass-star-forming complex because of the spatial resolution and sensitivity limitations of the HIFI instrument on the \textit{Herschel} Space Observatory. This latter survey, one of the most complete, only consists of single-pointing observations. Therefore, new instruments are needed to fully estimate the amount of H$_2$O emission coming from a forming Galactic cluster. 

To take another approach, we estimate the amount of H$_2$O emission from the nearby W3 high-mass-star-forming region. Its distance is 2 kpc and its age is 2 Myr \citep{bik2012}. We used a mass of 4$\times$10$^5$ M$_\odot$ for the entire cluster \citep{riveraingraham2013}, corresponding to a total luminosity of 2$\times$10$^6$ L$_\odot$ using Eq. \ref{eq:lum-to-mass}. To estimate the missing emission from all protostars, we ran a model for just one cluster instead of an entire galaxy. The cluster model predicts a total line intensity of 120 K km s$^{-1}$, which may be compared to the observed value of the high-mass protostar W3-IRS5 of 21.9 K km s$^{-1}$ \citep{vandertak2013}, which has a luminosity of 10$^5$ L$_\odot$, or 5\% of that of the cluster. The simulated value is highly sensitive to the adopted age of the cluster, for example, 1 Myr would result in a predicted intensity of 250 K km s$^{-1}$. This implies that for an individual cluster, we need to know the age accurately to within 10\%, which is not currently possible. It is reasonably possible that the amount of water emission we are missing is between a factor of 6 and 12. Without being able to map the entire cluster in water emission, we will not know exactly how much. 

\subsection{\texorpdfstring{High-$z$ test}{High-z test}}
\label{subsect:test}

Knowing that the relation between water emission and $L_\mathrm{FIR}$ spans over many orders of magnitude starting from the low-mass protostars to high-$z$ HyLIRGs, we probed the influence of the extragalactic observations on the $M_\mathrm{env} - I$ relation, and explore how this extrapolated form of the formula impacts the derived intensities. 

In Fig. \ref{fig:gal_highz}, we see that by including the extragalactic observations, we effectively lower the contribution from the low-mass end of the correlation and we see that it will only positively impact the high-mass protostars. On the other hand, the purely Galactic correlation lowers the emission from the high-mass protostars. Therefore, considering that we are underestimating water emission, we focused only on the standard and top-heavy IMF forms. We did this because the standard IMF is already dominated by the low-mass end of the distribution, and we also know from \cite{dutkowska2022} that the emission derived for the bottom-heavy IMF is practically indistinguishable from the standard one. At the same time, the top-heavy IMF would increase the emission even for the normal form of the correlation, and the inclusion of the extragalactic sources increases the slope by $\sim~10\%$ (see Fig. \ref{fig:gal_highz}). 

The results of the test indicate that inclusion of the extragalactic sources results in lowered emission,
on average, and that the difference with the results with the purely Galactic correlation starts to diminish for higher galactic masses and higher star-formation efficiencies. This effect is not surprising as the star-formation process is dominated in both total mass and number by low-mass protostars, while in terms of total bolometric luminosity the high-mass stars completely dominate the picture \citep[e.g.,][]{kroupa2002}. Therefore, the inclusion of the extragalactic sources, which lowers the emission from the low-mass protostars, naturally lowers the water emission derived from the simulated galaxies, as this is the main star-forming component if we consider Milky Way-like star formation. However, for the high-$z$ starbursts with high star-formation efficiencies and seemingly top- or even extremely top-heavy IMFs, this extrapolation could make a difference, when simulating star formation and its emission. Nevertheless, we do not investigate this further, as this is beyond the scope of this paper.

\subsection{SFR estimates}
\label{subsect:sfr_est}

From the results derived in this study, we are consistently overestimating SFRs for given galactic types. However, when considering the assumptions behind the model, and the fact that in the current version of the model, current SFRs are simulated without correcting for star formation histories or existing populations, the overestimation is no longer prominent.

The galaxy-in-a-box model was created as a tool for simulating emission from active and current star formation in galaxies. Therefore, even though the model accounts for dynamical differentiation of (proto)stellar ages, the model does not account for already existing, older stellar populations that normally would contribute to observations from which the rates are calculated. Moreover, as seen in Figs. \ref{fig:mass_sfr} and \ref{fig:lfir_sfr}, the results lie close to the literature estimates, if we assume low star formation activity. A calibration of the SFRs of galaxies depends on their current star formation
activity. If the bulk of galaxies are observed during a period of low star formation, we would naturally fall on the lower SFR side. Also, there are many factors influencing star formation activity in galaxies that  are not taken into account in the current version of the galaxy-in-a-box model. 

Another important aspect is that when calibrating SFRs from $L_\mathrm{FIR}$, one has to make assumptions about parameters such as the IMF and star formation history, which are sources of additional uncertainty in the final estimation of the SFR. Moreover, $L_\mathrm{FIR}$ is likely to underestimate the SFR in young clusters \citep{gutermuth2011} by up to an order of magnitude, and these are the main objects of interest in this study. If this is the case, our SFR estimates are roughly consistent with expectations. 

Lastly the galaxy-in-a-box model accounts for all stellar products, from brown dwarfs to high-mass stars. Therefore, it is not subject to observational limitations and the apparent overestimation could be an effect of accounting for all objects, including those that are normally unobservable, as illustrated in the W3 example above. The scenario we are considering slightly more closely resembles the high-$z$ situation, where galaxies are filled with active star-forming regions and are described as `full of Orions' \citep{rybak2020}. In this case, having relatively young star-forming regions, we trace only active and current star formation without accounting for higher differentiation of ages and stellar populations.

\subsection{Impact of the star-formation parameters}
\label{subsect:params}

In this study, we explored simulations for different galaxy types, and as such explored a broad parameter space (see Sect. \ref{subsect:parameters} and Table \ref{tab:params}). As in the first galaxy-in-a-box study \citep{dutkowska2022}, we observe no strong effect of the IMF, even though we included nearby subLIRGs, LIRGs, and quasars, as well as high-z quasars, ULIRGS, and HyLIRGs in the correlation that is used to assign molecular emission to protostars. This is expected as the extrapolation to the high-$z$ regime changes the slope of the correlation only by $\sim$10\%. 

We observe a strong impact of the star-formation efficiency and the free-fall-time scaling factor, both for the derived emission and SFRs. This is of no surprise as both parameters impact the stellar population of each cluster. The free-fall-time scaling factor will effectively lower the ages of the clouds and thus increase the emission, while the star-formation efficiency regulates how much of the molecular reservoir will be turned into stars, hence increasing the number of stars. 

One of the new input parameters in the galaxy-in-a-box model is the mass of the galaxy, as derived from Eq. (\ref{eq:lum-to-mass}). Clearly, the more massive the galaxy, the more emission we derive from the model. However, this parameter has its own uncertainty, which would be especially important  when considering the predicted water emission. The relation between the mass and luminosity was also derived for young stellar objects by \cite{pitts2022}, where:
\begin{equation}
    \log \left( M_\mathrm{env} / \mathrm{M}_\odot \right) = 0.30\substack{+0.07 \\ -0.06} + {0.79\substack{+0.01 \\ -0.02}} \log \left( L_\mathrm{bol} / \mathrm{L}_\odot \right) .\
\end{equation}
Although this expression was inferred for individual protostellar envelopes, it clearly agrees with Eq. \ref{eq:lum-to-mass} within the uncertainty. Here, we make the assumption that $\mathrm{L}_\mathrm{bol}$ represents $\mathrm{L}_\mathrm{FIR}$ as young protostars are deeply embedded in gas and dust, and $\mathrm{L}_\mathrm{bol}$ will be dominated by the contribution from $\mathrm{L}_\mathrm{FIR}$. Hence, if the relation between mass and luminosity is more universal, underestimating or overestimating can respectively underestimate or overestimate the available molecular reservoir.

\subsection{Comparison with observations}
\label{subsect:obs}

When comparing the derived values with observations, we clearly see that we are underestimating the water emission by at least one to two orders of magnitude (see Fig. \ref{fig:lfir_lh2o}) and overestimating the SFRs from a factor of a few to two orders of magnitude (see Fig.  \ref{fig:mass_sfr} and \ref{fig:lfir_sfr}). We discuss the possible explanations for the difference in SFR in Sect. \ref{subsect:sfr_est} extensively, and here we focus solely on the difference between our estimate and that of \cite{jarugula2019}. The SFR calibration of \cite{jarugula2019} utilizes the L$_\mathrm{FIR}$ -- SFR relation from \cite{kennicutt2012}:
\begin{equation}
    SFR\;(\mathrm{M}_\odot\,\mathrm{yr}^{-1}) = 1.47\times10^{-10} L_\mathrm{IR}\,(\mathrm{L}_\odot) ,\
\end{equation}
which, as mentioned in Sect. \ref{subsect:sfr_est}, is subject to various uncertainties. This is especially important when considering the IMF in the high-$z$ ULIRGs and HyLIRGs, as found in many studies \citep[e.g.,][]{zhang2018}, adding uncertainty to the calibration. Moreover if we were to apply the calibration from \cite{jarugula2019}, we would heavily underestimate SFRs towards well-studied, resolved Galactic clouds, where the relation inferred for water emission and luminosity is $\approx3000$ times higher than that of \cite{jarugula2019} \cite[further discussion in][]{kristensen2022sub}.

Focusing on the water emission, there are a few factors that could contribute to the observed difference and we discussed some of them in Sect. \ref{subsect:ratio}. Additionally, one of the reasons for not recovering the emission is that we do not convert 100\% of the galactic mass to an emitting source. There is a number of parameters standing in the way, with the star formation efficiency being the most obvious one. Moreover, currently we consider emission only from Class O and Class I protostars. Therefore, when considering emitting components that constitute only a small percentage of a whole galaxy, we are naturally going to lose a certain amount of emission. 

In galaxies there are more emitting components than simply protostars. These include photodissociation regions, galactic outflows, and supernovae. Even though their contribution is likely to be lower than that from star formation, their inclusion in calculations is essential in order to fully reproduce the emission, and as such, is a part of planned future improvements.

\section{Conclusions}
\label{sect:conclusions}

We extended the galaxy-in-a-box model to relate the predicted molecular emission from forming stars with SFRs. In this paper, we demonstrate the introduced extension and evaluate the derived results for galaxies with $L_\mathrm{FIR} = 10^8 - 10^{11} \mathrm{L}_\odot$ and various levels of star formation activity. We complemented the SFR study by extracting predicted emission for the para-H$_2$O \(2_{02} - 1_{11}\) line at 987.927 GHz. Our main results are as follows:
\begin{itemize}
    \item The star formation efficiency and the free-fall-time efficiency have a strong impact on the SFR and emission, whereas the opposite holds for the IMF.
    \item For the most extreme star-forming cases, the galaxy-in-a-box model overestimates the SFRs by up to two orders of magnitude. However, this difference could be lowered depending on the extent to which the current calibrations using $\mathrm{L}_\mathrm{FIR}$ as a star formation tracer underestimate the actual SFR values.
    \item The model underestimates the water emission by up to two orders of magnitude, and especially for the
high-$z$ quasars, ULIRGs, and HyLIRGs.
    \item For the moment, the model does not account for additional sources of emission, including supernovae, photodissociation regions, and galactic outflows. Moreover, we need to revisit the derived water emission for Galactic high-mass-star-forming regions, as we might miss the bulk of emission.
\end{itemize}

Our estimates deviate from observations and the literature. However, the apparent differences are consistent with expectations in the sense that known sources of emission are not included in the model, and therefore the galaxy-in-a-box model is a promising step toward shedding light on the star-forming properties of galaxies across cosmic time. In the near future, we plan to introduce a number of extensions that will account for other sources and processes that could contribute to the emission. The planned extensions include accounting for galactic outflows ---both AGN and starburst driven---, shocks from supernovae, and emission from photodissociation regions. Moreover, we are introducing H$_2$ and high-$J$ CO emission, which is going to be especially important in the JWST era. 

To properly account for water emission in our own Galaxy in the future, we will need a new far-infrared probe with the sensitivity of JWST. Such a probe is the planned PRIMA\footnote{\url{https://prima.ipac.caltech.edu}} mission. Only then will we be able to fully recover the emission from star-forming clusters in the Galaxy, and properly estimate the contribution from protostars in all stellar mass ranges.

\begin{table*}[]
\caption{Simulation results for the SFR -- $L_\mathrm{FIR}$ and $L_{\mathrm{H}_2\mathrm{O}}$ -- $L_\mathrm{FIR}$ relation}
\label{tab:relations} 
\centering
\begin{tabular}{Sccc|Sc|c}
\hline \hline
\multicolumn{3}{Sc|}{\multirow{2}{*}{Galactic type}} & \multicolumn{2}{Sc}{\multirow{2}{*}{\begin{tabular}[c]{@{}c@{}}Model results\\ $\mathrm{log}_{10}Y=a\cdot\mathrm{log}_{10}X+b$\end{tabular}}} \\
\multicolumn{3}{c|}{} & \multicolumn{2}{c}{}                           \\ \hline
$\varepsilon_\mathrm{SF}$  &  
$\tau^\mathrm{sc}_\mathrm{ff}$  & 
IMF &
$Y = \dfrac{SFR}{\mathrm{M}_\odot\,\mathrm{yr}^{-1}},\;X=\dfrac{L_\mathrm{FIR}}{\mathrm{L}_\odot}$  & 
$Y = \dfrac{L_{\mathrm{H}_2\mathrm{O}}}{\mathrm{L}_\odot},\;X=\dfrac{L_\mathrm{FIR}}{\mathrm{L}_\odot}$  \\ \hline

$1\%$  &  1  &   s   &  $a=0.908\pm0.021,\;b=-7.462\pm0.218;\; R^2=0.941$  &  $a=0.809\pm0.001,\;b=-6.685\pm0.013$ \\
$1\%$  &  5  &   s   &  $a=0.945\pm0.033,\;b=-8.586\pm0.329;\; R^2=0.772$  &  $a=0.806\pm0.002,\;b=-7.202\pm0.026$ \\
$1\%$  &  1  &  t-h  &  $a=0.923\pm0.030,\;b=-7.523\pm0.304;\; R^2=0.827$  &  $a=0.808\pm0.001,\;b=-6.606\pm0.010$ \\
$1\%$  &  5  &  t-h  &  $a=0.841\pm0.028,\;b=-7.424\pm0.294;\; R^2=0.849$  &  $a=0.805\pm0.002,\;b=-7.111\pm0.026$ \\
$1\%$  &  1  &  b-h  &  $a=0.903\pm0.030,\;b=-7.410\pm0.298;\; R^2=0.765$  &  $a=0.809\pm0.001,\;b=-6.707\pm0.010$ \\
$1\%$  &  5  &  b-h  &  $a=0.935\pm0.035,\;b=-8.499\pm0.354;\; R^2=0.879$  &  $a=0.810\pm0.003,\;b=-7.269\pm0.029$ \\
$10\%$ &  1  &   s   &  $a=0.900\pm0.021,\;b=-6.389\pm0.214;\; R^2=0.832$  &  $a=0.808\pm0.001,\;b=-5.675\pm0.014$ \\
$10\%$ &  5  &   s   &  $a=0.998\pm0.029,\;b=-8.160\pm0.287;\; R^2=0.870$  &  $a=0.812\pm0.003,\;b=-6.261\pm0.028$ \\
$10\%$ &  1  &  t-h  &  $a=0.908\pm0.019,\;b=-6.348\pm0.189;\; R^2=0.982$  &  $a=0.811\pm0.001,\;b=-5.632\pm0.010$ \\
$10\%$ &  5  &  t-h  &  $a=0.925\pm0.027,\;b=-7.293\pm0.272;\; R^2=0.908$  &  $a=0.810\pm0.002,\;b=-6.157\pm0.019$ \\
$10\%$ &  1  &  b-h  &  $a=0.892\pm0.024,\;b=-6.310\pm0.238;\; R^2=0.956$  &  $a=0.807\pm0.002,\;b=-5.690\pm0.017$ \\
$10\%$ &  5  &  b-h  &  $a=0.934\pm0.024,\;b=-7.521\pm0.247;\; R^2=0.948$  &  $a=0.812\pm0.003,\;b=-6.295\pm0.028$ \\
$30\%$ &  1  &   s   &  $a=0.918\pm0.066,\;b=-6.079\pm0.624;\; R^2=0.658$  &  $a=0.807\pm0.001,\;b=-5.182\pm0.013$ \\
$30\%$ &  5  &   s   &  $a=0.941\pm0.031,\;b=-7.083\pm0.297;\; R^2=0.892$  &  $a=0.813\pm0.002,\;b=-5.795\pm0.025$ \\
$30\%$ &  1  &  t-h  &  $a=0.899\pm0.034,\;b=-5.751\pm0.330;\; R^2=0.832$  &  $a=0.809\pm0.001,\;b=-5.135\pm0.012$ \\
$30\%$ &  5  &  t-h  &  $a=0.865\pm0.025,\;b=-6.167\pm0.240;\; R^2=0.923$  &  $a=0.805\pm0.003,\;b=-5.620\pm0.028$ \\
$30\%$ &  1  &  b-h  &  $a=0.870\pm0.032,\;b=-5.611\pm0.310;\; R^2=0.789$  &  $a=0.810\pm0.002,\;b=-5.244\pm0.016$ \\
$30\%$ &  5  &  b-h  &  $a=0.873\pm0.038,\;b=-6.354\pm0.371;\; R^2=0.861$  &  $a=0.808\pm0.003,\;b=-5.771\pm0.032$ \\
\hline
$1\%$  &  1  &   s   &  ---  &  $a=0.809\pm0.001,\;b=-5.668\pm0.008$   \\
$1\%$  &  1  &  t-h  &  ---  &  $a=0.808\pm0.001,\;b=-6.655\pm0.012$   \\
$10\%$ &  1  &   s   &  ---  &  $a=0.811\pm0.002,\;b=-5.806\pm0.016$   \\
$10\%$ &  1  &  t-h  &  ---  &  $a=0.810\pm0.001,\;b=-5.677\pm0.010$   \\ \hline

\end{tabular}
\tablefoot{Results from running simulations with all considered parameters combinations (for more details see Tab. \ref{tab:params}). In the IMF parameters, \textit{s} refers to standard, \textit{t-h} to top-heavy, and \textit{b-h} to bottom-heavy. For details on the $\tau^\mathrm{sc}_\mathrm{ff}$ and its relation with the free-fall time efficiency we refer the reader to \cite{dutkowska2022}, where extensively discussed the impact of this factor. The four setups in the bottom of the table divided by the horizontal line refer to our high-$z$ test (see. Sect. \ref{subsect:test}). Since the test had impact only on the water emission, we did not provide the SFR -- $L_\mathrm{FIR}$ relations as these are the same as for the corresponding Galactic setups. The correlation between the $L_{\mathrm{H}_2\mathrm{O}}$ -- $L_\mathrm{FIR}$ had consistent $R^2 \approx 0.999$, hence we do not provide the $R^2$ values for this relation in the table.}
\end{table*}

\begin{acknowledgements}
     The research of KMD and LEK is supported by a research grant (19127) from VILLUM FONDEN.
\end{acknowledgements}
\bibliographystyle{aa}
\bibliography{refs}

\begin{appendix}

\section{Overview of the galaxy-in-a-box model}
\label{app:giab-overview}

The galaxy-in-a-box model, which was used to derive the results presented in this study, is described in detail in the paper by \cite{dutkowska2022}. Below we provide a general description.

The architecture of the galaxy-in-a-box model is rooted in three elements of the galactic star-forming environment, i.e., giant molecular clouds (GMCs), star-forming clusters, and protostars. The model starts the calculations by generating a spatial and mass distribution of GMCs based on the observational data of other galaxies. The GMCs to be passed to the next steps of the simulation are chosen randomly from the mass distribution, and each GMC mass acts as an initial cluster mass, i.e., one GMC will form one cluster in the model. Before the GMC mass is passed to the cluster module \citep[based on the cluster-in-a-box model by][]{kristensen2015}, each cloud is assigned an age based on its free-fall time, which is then randomly scaled between being newly formed and completely collapsed. This affects the number of deeply embedded protostars (Class 0 and I protostars) driving outflows in a given cloud. In the next step of the calculations, the cluster module returns protostellar mass, age, and spatial distribution. However, the latter is disregarded in the galactic-scale calculations. 

With the complete protostellar distribution, molecular emission is assigned to each Class 0 and Class I protostar. This results in the total expected molecular outflow emission from a given cluster. The model repeats these calculations for all chosen GMCs. The information about the total cluster mass (expressed as the total mass of (proto)stellar content) and emission is then returned to the galactic spatial grid. After accounting for the sizes of clusters and their location, the raw galactic emission image is convolved with a Gaussian beam. Hence, the model returns statistics on galactic clusters (their number of stars, mass, and emission) and an integrated intensity image. However, the level of detail in the returned statistics can be easily adjusted, such that the exact protostellar distributions of each cluster can be stored.

\end{appendix}

\end{document}